\documentclass[journal]{IEEEtran}

\usepackage{graphicx}
\DeclareGraphicsExtensions{.pdf,.jpeg,.jpg,.png}

\ifCLASSOPTIONcompsoc
    \usepackage[caption=false, font=normalsize, labelfont=sf, textfont=sf]{subfig}
\else
\usepackage[caption=false, font=footnotesize]{subfig}
\fi

\usepackage{amsmath}
\usepackage{amssymb}
\usepackage{physics}
\usepackage{xcolor}
\usepackage[super]{nth}
\usepackage{algorithm}
\usepackage{algpseudocode}

\usepackage{cases}
\usepackage{cite}

\newcommand{\etal}{\textit{et al}. }

\begin{document}
\title{LASSO-Based Multiple-Line Outage Identification In Partially Observable Power Systems}

\author{
Xiaozhou~Yang%
~and~Nan~Chen,~\IEEEmembership{Member,~IEEE}

\thanks{X. Yang is with the Singapore-ETH Centre, Future Resilient Systems, CREATE campus, 1 CREATE Way, \#06-01 CREATE Tower, Singapore 138602, Singapore. 
Nan Chen is with the Department of Industrial Systems Engineering and Management, National University of Singapore, Singapore 117576, Singapore.
To whom correspondence may be addressed: isecn@nus.edu.sg}%
}

\maketitle

\begin{abstract}
Phasor measurement units (PMUs) create ample real-time monitoring opportunities for modern power systems. Among them, line outage detection and identification remains a crucial but challenging task. 
Current works on outage identification succeed in full PMU deployment and single-line outages. Performance however degrades for multiple-line outage with partial system observability. 
We propose a novel framework of multiple-line outage identification using partial nodal voltage measurements. Using alternating current (AC) power flow model, phase angle signatures of outages are extracted and used to group lines into minimal diagnosable clusters. Identification is then formulated into an underdetermined sparse regression problem solved by lasso. Tested on IEEE 39-bus system with 25\% and 50\% PMU coverage, the proposed identification method is 93\% and 80\% accurate for single- and double-line outages. 
Our study suggests that the AC power flow is better at capturing outage patterns and sacrificing some precision could yield substantial improvement in identification accuracy. These findings could contribute to the development of future control schemes that help power systems resist and recover from outage disruptions in real time. 

\end{abstract}


\section{Introduction}
\IEEEPARstart{C}{ritical} infrastructure resilience is an emerging research topic for which the importance is increasingly recognized among researchers, practitioners, and policy makers. 
Power system is one of the most critical infrastructures for a normally functioning society. As part of the global drive for power grid modernization, tools for real-time situational awareness in the control room, e.g., knowledge about system contingencies, are being developed. 
In particular, power system transmission line outage is one type of disruption that attracts significant attention due to its frequent occurrence and tremendous damage if not addressed in time  \cite{patel2010real}. To facilitate the recovery phase of a resilient system, a disruption must be detected and accurately located. Over the years, various schemes have been developed to detect outages as quickly as possible, e.g., see \cite{yang2020control,yang2021particle}. However, their focus is on fast detection rather than accurate location identification. This work deals with the problem of identifying the true outage lines when a detection alarm has been raised. 

There are two main challenges to accurate outage identification. The first is limited observability. Phasor measurement unit (PMU), a sensor that samples bus voltage phasors at high frequency and with GPS time synchronization, is the key to many real-time power system monitoring capabilities. However, due to the progressive development and an overwhelming set of real-time data, one must consider a limited PMU deployment in the system when designing an outage identification scheme \cite{aminifar2014synchrophasor}. The proposed scheme therefore has to maximize its performance under the constraint of limited observability. The second challenge is the scalability of the scheme give the inherent combinatorial nature of potential outage locations. An outage can happen at one or multiple transmission lines; the number is in general not known a priori. For example, for a system with $L$ transmission lines, the solution space consists of $2^L$ outage location combinations. Any proposed scheme has to overcome this challenge to realize fast and accurate system-wide outage identification.

\section{Related Works in Outage Identification}
Power system line outage localization is in general a line parameter change identification problem. An outage corresponds to the change when the line admittance drops to zero. Some attempted to solve the general problem by recovering changed line parameter. Yu \etal uses the power flow equations to formulate line parameters as unknown regression coefficients and recover them via least squares method \cite{Yu2018}. Another approach is to focus on the system admittance matrix and recover the elements of the matrix via matrix decomposition \cite{Ardakanian2019a} or lasso \cite{Babakmehr2016}. These methods can both locate outages and recovery parameters. But they require full PMU and/or smart meter measurements, e.g., $P, Q, V$ for the power flow approach or $I, V$ for the admittance matrix approach at all buses. 

It is however more realistic to assume only part of the system is observable by PMUs or smart meters. Operating under this assumption, research works mainly seek to solve the line outage localization problem, and they generally take two approaches. Machine learning-based approach has been gaining traction in recent years. This line of work leverages easily accessible simulated outage data and trains an outage classifier to identify the most likely outage location. For example, Garcia \etal, Kim and Wright train multinomial regression classifiers, a type of supervised classification algorithm, while Li \etal and Zhao \etal train neural networks, another versatile machine learning framework, to identify tripped lines \cite{Garcia2016,Kim2018,Li2019a,Zhao2020}. This approach is powerful at locating multiple line outages with limited PMUs. However, their performance depend on generalizable training data. 

The other approach that does not depend on training data and considers partial observability is the expected nodal voltage angle change formulation. Based on direct current (DC) \cite{Tate2008,Zhu2012,Chen2014,Wu2015} or AC \cite{Costilla-Enriquez2019} power flow models, a dictionary of angle changes is built in advance. Identification is then formulated into an unknown sparse vector recovery problem solved by greedy optimization methods \cite{Zhu2012, Chen2014} and matrix decomposition \cite{Wu2015}, or a pattern-matching problem solved by correlation-based methods \cite{Tate2008,Costilla-Enriquez2019}. 
This approach leverages physical laws governing power systems and integrate them with data-driven methods. This is the closest line of work to ours. However, the assumption of DC model potentially reduces system state fidelity critical for accurate identification. Despite demonstrated effectiveness by the above works, the identification performance degrades significantly when fewer PMUs are available or when multiple-line outages are considered. 

Therefore, given an outage detection timestamp, multiple-line outage identification problem still remains difficult when 1) massive generalizable training data is not available or feasible, 2) only a portion of the system buses are equipped with PMUs, 3) only nodal voltage phasor information is used.  
To address these gaps, we propose a new framework of multiple-line outage identification based on power system sensitivity analysis and sparse regression methods considering line diagnosabilities. Using readily available system topology and parameter information, we build a signature map of line outages based on AC power flow sensitivity analysis. Outage identification problem is then formulated into an underdetermined sparse regression problem that accommodates any a priori unknown number of simultaneous line outages. Crucially, clusters of lines whose outages are indistinguishable under the given PMU placement are identified and augmented with the initial result to improve identification accuracy.  

Our contributions can be summarized in three aspects: 1) We improve the state-of-the-art multiple-line outage identification performance under limited PMU deployment; 2) The novel sparse regression formulation accommodates unknown number of outage lines and is robust with noisy data; 3)
We also propose a way to account for indistinguishable outages using minimal diagnosable clusters which significantly improve overall identification accuracy.

The rest of this paper is organized as follows. The basis for outage identification is the post-outage voltage angle signature and is derived in Section \ref{sec:angle_signature}. Multiple-line outage identification scheme is then described in Section \ref{sec:identification_scheme}. Section \ref{sec:simulation} demonstrates the effectiveness of the proposed scheme compared to existing ones. We conclude this work and discuss future research directions in Section \ref{sec:conclusion}. 

\section{Phase Angle Signature of Outages}
\label{sec:angle_signature}

Each outage is different. Machine learning-based approaches let algorithms learn the difference through clever training and generalizable data. Physics-informed approaches, e.g., our proposed method, leverage physical laws governing power systems to find out the difference instead. In general, two questions need to be answered to build an effective physics-informed outage identification scheme: 1) How to quantify the impact of each line outage to nodal bus state variables? 2) Given the characterization, how to identify the most probable outage lines out of all possible ones? We explain in this section how the first question can be answered through sensitive analysis on AC power flow model. In the next section, an efficient and robust identification scheme is developed. 

\subsection{Power Flow Model}
Consider a power system with $N$ buses and $L$ transmission lines where $\mathcal{N} = \{1, 2, \dots, N\}$ and $\mathcal{L} = \{1, 2, \dots, L\}$. AC power flow model is the governing equation between active and reactive power injection (P, Q) and voltage phasor (V$\angle\theta$) at each bus \cite{Glover2012}:
\begin{subequations}
\label{eqn:AC_power_flow_model}
\begin{align}
\text{P}_m &= \text{V}_m \sum_{n=1}^{N} \text{V}_n \text{Y}_{mn} \cos (\theta_m - \theta_n - \alpha_{mn}) \,, \label{eqn:AC_power_flow_P}\\
\text{Q}_m &= \text{V}_m \sum_{n=1}^{N} \text{V}_n \text{Y}_{mn} \sin (\theta_m - \theta_n - \alpha_{mn}) \,, \label{eqn:AC_power_flow_Q}
\end{align}
\end{subequations}
where $m \in \mathcal{N}$, and $\boldsymbol{Y}$, the bus admittance matrix of which Y$_{mn}\angle\alpha_{mn}$ is the $(m,n)_{th}$ element, is assumed to be known. $\text{V}_m$ and $\theta_m$ are also assumed to be available if bus $m$ has a PMU. Let \textbf{P}, \textbf{Q}, $\boldsymbol{\theta}$, and \textbf{V} represent the $N$-vectors of active and reactive power injection, voltage angles, and magnitudes at all buses except the reference bus\footnote{By convention, bus 1 is assumed to be the reference bus whose voltage phase angle is set to $0^\circ$ and magnitude to $1.0$ per unit (p.u.).}. 
A sensitivity analysis on power injections by linearization of (\ref{eqn:AC_power_flow_P}) around a pre-outage steady-state operating point yields the following partial differential equation: 
\begin{equation}
\label{eqn:ac_decoupled_jacobian}
\Delta \textbf{P} \approx J_1 \Delta\boldsymbol{\theta} + J_2 \Delta\textbf{V}\,,
\end{equation}
where ${J}_1, {J}_2$ are two submatrices of the AC power flow Jacobian with 
\begin{equation}
\label{eqn:ac_jacobian}
{J}_1 = \frac{\partial \textbf{P}}{\partial \boldsymbol{\theta}} \,, 
{J}_2 = \frac{\partial \textbf{P}}{\partial \textbf{V}} \,.
\end{equation}
We let $\Delta \textbf{P} = \textbf{P}' - \textbf{P}$ where $\textbf{P}$ and $\textbf{P}'$ denote pre- and post-outage bus power injections. $\Delta\boldsymbol{\theta} = \boldsymbol{\theta}' - \boldsymbol{\theta}$ is similarly defined. In the usual operating range of relatively small angles, power systems exhibit much stronger interdependence between \textbf{P} and $\boldsymbol{\theta}$ as compared to \textbf{P} and \textbf{V} \cite{murty2017power}. Therefore, we focus on the relationship between real power injection and voltage phase angle, i.e., ${J}_1$, in the remainder of the paper\footnote{The same set of analysis can applied to reactive power and voltage magnitude as well, which we omit here. We also skip some details about power system linearization as the formulation is standard. Interested reader can refer to Section II of \cite{yang2020control}.}. Redefining ${J}_1$ as ${J}$, the off-diagonal and diagonal elements of ${J}$ can be derived from (\ref{eqn:AC_power_flow_P}) as:
\begin{subequations}
\label{eqn:elements_J}
\begin{align}
    \frac{\partial \text{P}_{m}}{\partial \theta_{n}} 
    & = \text{V}_{m} \text{V}_{n} \text{Y}_{m n} \sin \left( \theta_{m} - \theta_{n} - \alpha_{m n} \right) \,,  m \neq n \,,\label{eqn:elements_J_off}\\ 
    \frac{\partial \text{P}_{m}}{\partial \theta_{m}} 
    & = -\sum_{ n=1 \atop n \neq m}^{N} \text{V}_{m} \text{V}_{n} \text{Y}_{m n} \sin \left( \theta_{m} - \theta_{n} - \alpha_{m n} \right) \,. \label{eqn:elements_J_diag} 
\end{align}
\end{subequations}

\subsection{Outage Signature Map}
We established an AC power flow-based relationship between instantaneous real power injection changes and voltage phase angle changes as $\Delta\textbf{P} = {J} \Delta\boldsymbol{\theta}$. Inverting the Jacobian matrix, the relationship can be written as:
\begin{equation}
\label{eqn:angle_sensitivity}
\Delta\boldsymbol{\theta}  = {J}^{-1} \Delta\textbf{P} \,.
\end{equation}
Therefore, ${J}^{-1}$ in (\ref{eqn:angle_sensitivity}) quantifies the impact on bus angles associated with respect to a unit change in real power injections. Under the DC assumption, a neat way to characterize the impact of an outage at line $l$ carrying power from bus $i$ to $j$ is a real power injection of $p_{l}$ at bus $i$ and withdrawal of $-p_{l}$ at bus $j$ \cite{wood2013power}. Equivalently, the change in real power injection due to an outage at line $l$ can be written as:
$$
\Delta \textbf{P} = p_{l} \cdot m_l \,.
$$ 
$m_l$ is an $N$-vector of zeros except with $1$ at the $i_{th}$ and $-1$ at the $j_{th}$ position. $p_l$ is a constant that depends on the pre-outage power flow and the so-called power transfer distribution factor \cite{liu2004role}. In general, we can obtain the expected change in phase angles due to all outages at line $l_i, i \in \mathcal{L}$. Putting the $p_{l}$ and $m_l$ for all transmission lines together and write in matrix form:
\begin{align}
\label{eqn:complete_map}
    [\Delta\boldsymbol{\theta}] 
    &= {J}^{-1} 
    \begin{bmatrix} p_{l_1} m_{l_1} & p_{l_2} m_{l_2} & \cdots & p_{l_L} m_{l_L}\\ 
    \end{bmatrix} \nonumber\\ 
    &= {J}^{-1} 
    \begin{bmatrix} m_{l_1} & m_{l_2} & \cdots & m_{l_L} \\
    \end{bmatrix} \operatorname{ diag}(\boldsymbol{p}) \nonumber\\
    &= {J}^{-1} M \operatorname{ diag}(\boldsymbol{p}) \,, 
\end{align}
where $M$ is the $N \times L$ bus to branch incidence matrix with columns corresponding to lines and rows to buses. $\operatorname{diag}(\boldsymbol{p})$ is the diagonal matrix with individual line power transfer $p_l$ on the diagonal. $M$ encodes the baseline system topology information, i.e., which bus is connected by which group of lines. It is also closely related to the bus admittance matrix $\boldsymbol{Y}$ since 
\begin{equation}
\label{eqn:admittance_matrix}
\boldsymbol{Y} = M \operatorname{diag}(\mathbf{y}) M^{\top} \,,
\end{equation}
where ${M}^{\top}$ is the transpose of ${M}$ and $\mathbf{y}$ is the vector of individual line admittance. 

In a realistic setting of limited PMU deployment, we assume there are fewer PMUs than buses and transmission lines, i.e. $K \le N$ and $K \le L$. Define a bus selection matrix $S\in\{0,1\}^{K\times N}$ that selects rows of buses with PMUs, the observable phase angle impact from all line outages is
\begin{align}
\label{eqn:observable_impact}
[\Delta\boldsymbol{\theta}]_I &= S J^{-1} M \operatorname{diag}(\boldsymbol{p}) \nonumber\\
& = F \operatorname{diag}(\boldsymbol{p}) \,,
\end{align}
where we let $F = S J^{-1} M$. Therefore, $F$ is a $K \times L$ outage signature map determined by PMU locations, system operating states, and topology. Each column of $F$, i.e., $F_{l}$ represents the incremental effect of line $l$ outage on all bus voltage angles captured by PMUs. 

\begin{figure}[t]
\centering
\includegraphics[width=1\linewidth]{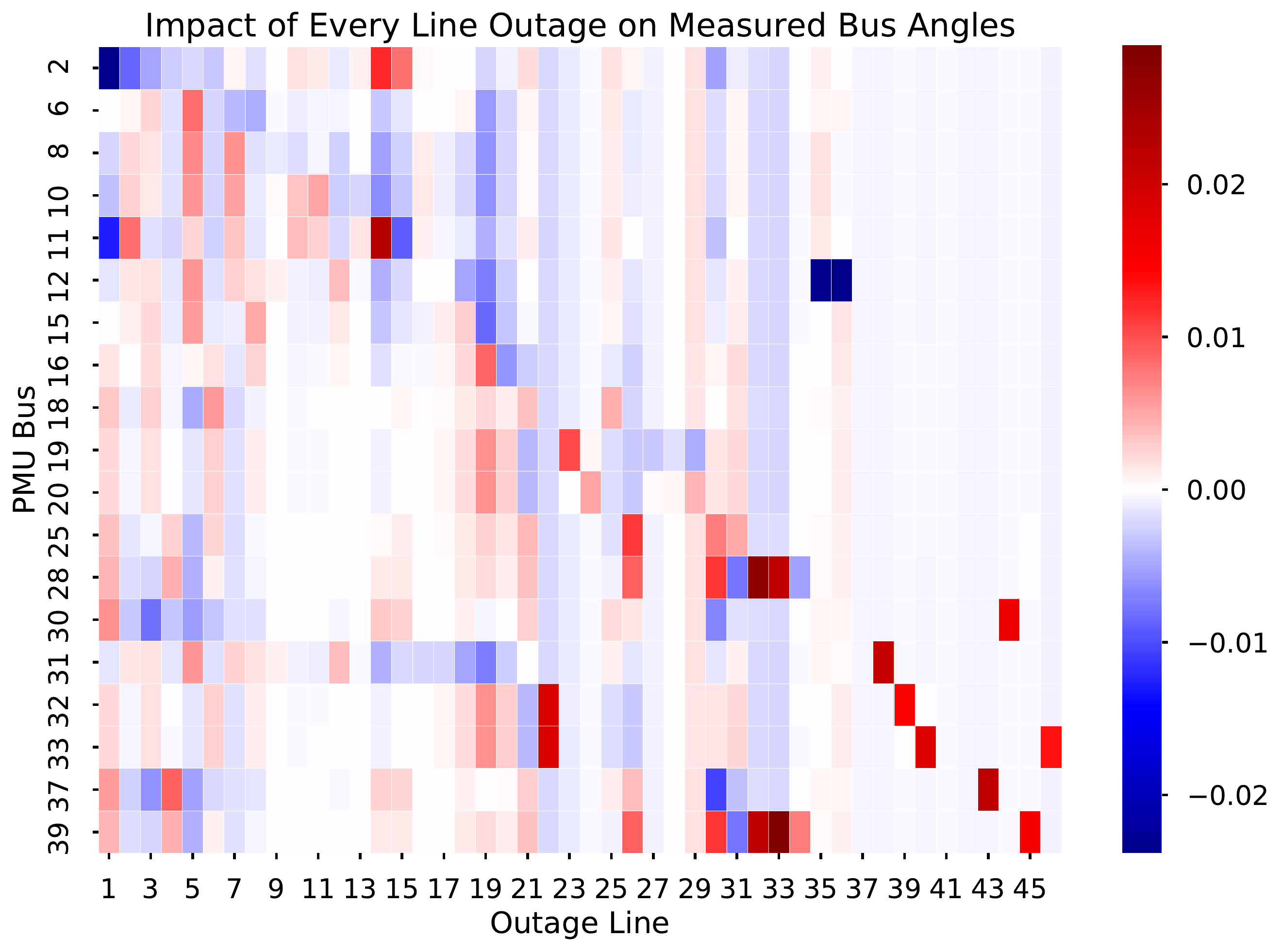}
\caption{\textit{An example of the $19 \times 46$ signature map constructed using a random placement of 19 PMUs in the New England 39-bus system with 46 transmission lines. Each column corresponds to a single line outage and its incremental impact on PMU-equipped bus voltage phase angles.}}
\label{fig:feature_map}
\end{figure}
Fig. \ref{fig:feature_map} shows an example of $F$ for a random placement of 19 PMUs on the New England 39-bus system, using $J^{-1}$ obtained from steady-state bus voltages. The signature map captures the varying degree of impact each line outage has on PMU-equipped buses. Outages of some lines might not be uniquely identified because they create similar phase angle responses, e.g., line 10 and 11, line 32 and 33. The map also suggests that some line outages create minimal impact that they might be indistinguishable from normal conditions, e.g., line 37 or line 41. While distinctive signatures from the map should be useful in identifying outage lines, we need to address the problem of indistinguishable line outages in order to fully exploit the signature map information.

\section{Outage Identification Scheme}
\label{sec:identification_scheme}
After a single- or multiple-line outage, the signature map developed in the previous section provide a basis for accurate outage identification. We assume that the outage event is detected quickly using a detection scheme, e.g., of \cite{yang2020control} or \cite{yang2021particle}. We first formulate multiple-line outage identification as a sparse regression problem. Then, a method to address indistinguishable outages is proposed to further improve the regression result accuracy. 

\subsection{Identification by Sparse Regression}
\label{sec:sparse_regression}
Suppose $a$ simultaneous line outages happen at $l_i, i=1, \dots, a$ and the size is relatively small, i.e., $a \ll L$. Let $\boldsymbol{\beta}$ be an $L$-vector with all zeros except at $\boldsymbol{\beta}_{l_i}$ with value $p_{l_i}$ for $i = 1, \dots, a$.
If the outage model in Section \ref{sec:angle_signature} holds, we can write
\begin{align}
    \Delta\boldsymbol{\theta} &= S J^{-1} (m_{l_1} p_{l_1} + \dots + m_{l_a} p_{l_a}) + \boldsymbol{\epsilon} \nonumber\\
    &= S J^{-1} M \boldsymbol{\beta} + \boldsymbol{\epsilon} \nonumber\\
    &= F \boldsymbol{\beta} + \boldsymbol{\epsilon}  \,,
\end{align}
where $M$ in the second step is as defined in (\ref{eqn:complete_map}) and $\boldsymbol{\epsilon}$ is a Gaussian noise term with mean zero and known variance $\sigma^2 \textbf{I}_{K\times K}$, representing measurement error of $K$ PMUs. Hence, non-zero entries, or support, of the power transfer vector $\boldsymbol{\beta}$ reveal true outage locations. Given the signature map $F$ and PMU measurements $\Delta\boldsymbol{\theta}$, $\boldsymbol{\beta}$ can be estimated from the above relationship by minimizing the squared-error loss, subject to the outage size constraint as
\begin{align}
\label{eqn:l_0_formulation}
\underset{\boldsymbol{\beta} \in \mathbb{R}^{L}}{\min} &\left\|\Delta\boldsymbol{\theta} - F\boldsymbol{\beta}\right\|_{2}^{2} \,, \\
& \text { s.t. }\|\boldsymbol{\beta}\|_{0} = a \nonumber\,,
\end{align}
where $\left\|\cdot\right\|^2_2$ is the square of the $\ell_2$ norm and $\left\|\cdot\right\|_0$ is the number of non-zero entries of a vector. However, as we do not known a priori the location of the non-zero entries, the above formulation presents a challenging combinatorial optimization problem. Methods such as exhaustive search, forward-stepwise regression or mix integer optimization could be used to solve the problem \cite{bertsimas2020sparse}. However, compared to shrinkage method, in particular lasso, they are computationally more intensive, thus not suitable for real-time application in realistic power systems \cite{hastie2020best}. 

Lasso was originally proposed in \cite{tibshirani1996regression} and has since been used in various applications for ease of implementation, robustness to noise, and the ability to shrink some coefficients to exactly zero, thus recovering true support of the vector. Lasso solves a relaxed version of the problem in (\ref{eqn:l_0_formulation}) by replacing the $\ell_0$ constraint with an $\ell_1$ constraint:
\begin{align}
\label{eqn:l_1_formulation}
\underset{\boldsymbol{\beta}  \in \mathbb{R}^{L} }{\min} &\left\|\Delta\boldsymbol{\theta} - F\boldsymbol{\beta}\right\|_{2}^{2} \,, \\
& \text { s.t. }\|\boldsymbol{\beta}\|_{1} \le a \nonumber\,,
\end{align}
and equivalently in Lagrangian form:
\begin{equation}
\label{eqn:lasso_formulation}
\min_{\boldsymbol{\beta} \in \mathbb{R}^{L}}\left\{\|\Delta\boldsymbol{\theta} - F\boldsymbol{\beta}\|_{2}^{2}+\lambda\|\boldsymbol{\beta}\|_{1}\right\},
\end{equation}
where $\|\cdot\|$ is the $\ell_1$ norm and $\lambda$ is a regularization parameter that has one-to-one correspondence to $a$ for solutions of (\ref{eqn:l_1_formulation}) and (\ref{eqn:lasso_formulation}). Larger values of $\lambda$ impose stronger regularization on $\boldsymbol{\beta}$, whereas if $\lambda = 0$, the lasso solution $\boldsymbol{\hat{\beta}}$ is the same as the least squares estimate. According to \cite{zou2007degrees}, given a fixed $F$, there exists a finite sequence,
\begin{equation}
\label{eqn:transition_points}
\lambda_0 > \lambda_1 > \cdots > \lambda_Q = 0 \,,
\end{equation}
such that 1) for all $\lambda > \lambda_0$, $\boldsymbol{\hat{\beta}} = \mathbf{0}$, 2) the support of $\boldsymbol{\hat{\beta}}$ does not change with $\lambda$ for $\lambda_q < \lambda < \lambda_{q+1}\,, q=0, \dots, Q-1$. These $\lambda_q$'s are called transition points as the support in lasso solution changes at each $\lambda_q$. Often, $\lambda$ is selected according to some parameter tuning scheme, e.g. cross validation, such that the resultant regression model achieves best prediction accuracy. However, our objective is to uncover the true support of $\boldsymbol{\beta}$. Thus, lasso solution at various transition points need to be obtained each time an outage is detected to ascertain the location, and in effect the number, of outage lines. 

Least angle regression (LARS), originally proposed by \cite{efron2004least}, with lasso modification is an efficient algorithm that computes the entire lasso path with a complexity of least squares regression. Briefly, starting with coefficients of zero, LARS identifies the first variable as the one most correlated with the response, e.g., $\Delta\boldsymbol{\theta}$. As the coefficients of the active variables move toward their least squares estimates, a new variable becomes active when its correlation with the residual ``catches up'' with the active set. These changes happen at the transitions points of (\ref{eqn:transition_points}) and variables enter one at a time \cite{efron2004least}. The process is stopped after $Q$ steps and in general $Q = \min\{K-1, L\}$ for standardized data unless otherwise specified. 

Assuming $\Delta\boldsymbol{\theta}$, its mean $\Delta\bar{\boldsymbol{\theta}}$, signature map $F$, and the maximum number of non-zero entries $Q$ are provided, LARS can produce a sequence of regularization parameters $\lambda^q$ and the associated lasso solution $\boldsymbol{\beta}^q$ to (\ref{eqn:lasso_formulation}) as described in Algorithm \ref{alg:lars}. Indices of siginificantly non-zero entries of $\boldsymbol{\beta}^Q$ are then identified as potential outage locations. 
\begin{algorithm}
\caption{Least Angle Regression with Lasso Modification}
\label{alg:lars}
Input: $\Delta\boldsymbol{\theta}, \Delta\boldsymbol{\bar{\theta}}, F, Q$\\
Output: Lasso solution path $\{\lambda_q, \boldsymbol{\beta}^q\}_{q=0}^{Q}$
\begin{algorithmic}[1]
\State Standardize columns of $F$ to mean zero and unit $\ell_2$ norm. 
Set $\boldsymbol{\beta}^0 = (\beta_1, \beta_2, \dots, \beta_L) = \mathbf{0}$. Let $r_0 = \Delta\boldsymbol{\theta} - \Delta\boldsymbol{\bar{\theta}}$. 

\State Get first active column: 
$$
j = \arg\underset{i \in \mathcal{L}}\max|\langle r_0, F_i\rangle | \,.
$$
Let $\lambda_0 = |\langle r_0, F_j\rangle |$. Define $\mathcal{A} = \{j\}$ and $F_{\mathcal{A}}$ as the active set and signature matrix with columns from the set. 

\For{$q = 1, 2, \dots, Q$}
\State Get current least-squares direction: 
$
\delta = \frac{1}{\lambda_{q - 1}}(F_{\mathcal{A}}^{\top}F_{\mathcal{A}})^{-1}F_{\mathcal{A}}^{\top}r_{q-1} \,.
$ 
Define $L$-vector $\mathbf{u}$ such that $\mathbf{u}_{\mathcal{A}} = \delta$ and zero everywhere else.

\State  Move coefficients toward least-squares estimate: $\boldsymbol{\beta}(\lambda) = \boldsymbol{\beta}^{q-1} + (\lambda_{q-1} - \lambda) \mathbf{u}$, for $0 < \lambda \le \lambda_{q-1}$ while maintaining $r(\lambda) = \Delta\boldsymbol{\theta} - F \boldsymbol{\beta}(\lambda)$. Drop any element of $\mathcal{A}$ if the corresponding coefficient crosses 0 and recompute the least-squares estimate. 

\State Identify the largest $\lambda$ at which $|\langle r(\lambda), F_l \rangle| = \lambda$ for $l \notin \mathcal{A}$. Let $\lambda_p = \lambda$, the new transition point.

\State Suppose the new active column has index $j$. Update $\mathcal{A} = \mathcal{A} \cup j$, $\boldsymbol{\beta}^q = \boldsymbol{\beta}(\lambda_q) + (\lambda_{q-1} - \lambda_q)\mathbf{u}$, and $r_q = \Delta\boldsymbol{\theta} - F \boldsymbol{\beta}^q$.
\EndFor

\State Return the sequence $\{\lambda_q, \boldsymbol{\beta}^q \}_0^Q$.
\end{algorithmic}
\end{algorithm}

\begin{figure}[t]
\centering
\includegraphics[width=1\linewidth]{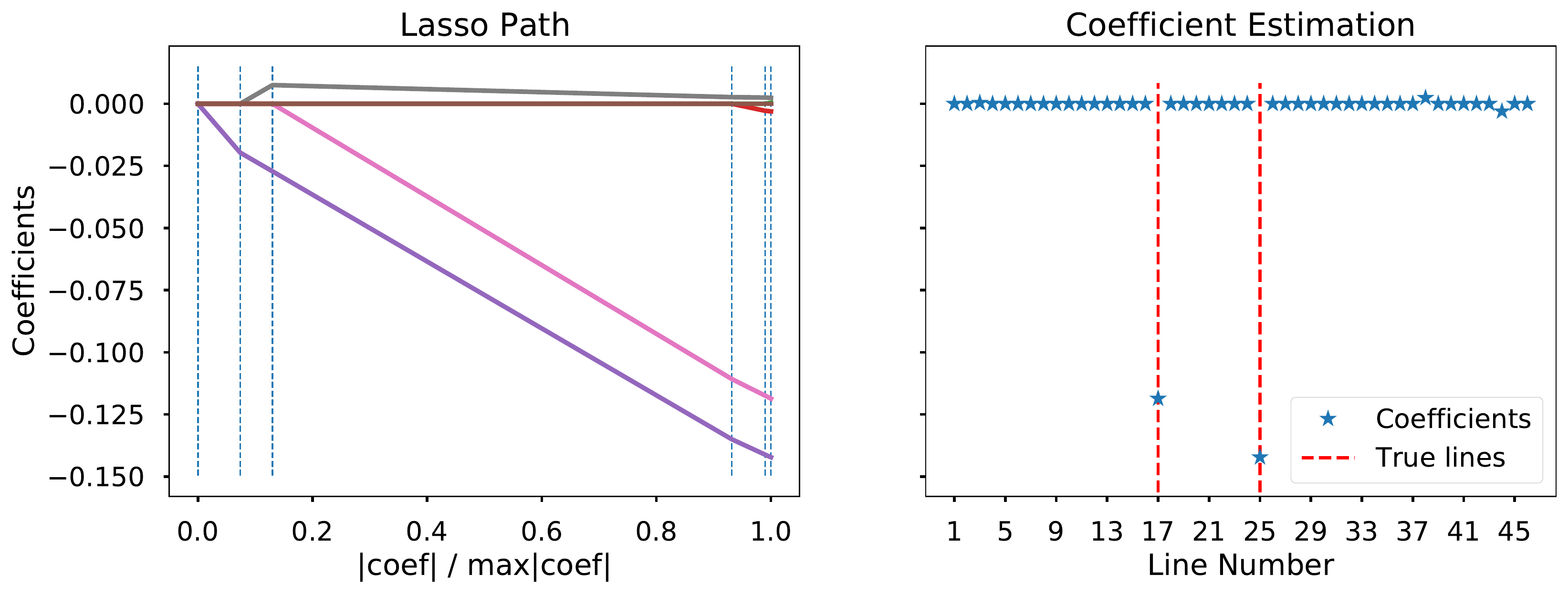}
\caption{\textit{Lasso path via LARS illustration for double-line outage at line 17 and 25. Complete lasso regularization path is shown on the left and coefficient estimation after five candidates entered the model on the right.}}
\label{fig:lasso_illustration}
\end{figure}
Fig. \ref{fig:lasso_illustration} shows an example of the lasso path computed using LARS for a double-line outage event. The final $\boldsymbol{\beta}$ has five non-zero coefficients after five transitions. The scheme correctly identifies line 17 and 25 as they have significantly non-zero estimated coefficients compared to the others. The third-highest coefficient corresponds to line 38 which is a neighbor of line 17 that likely produces similar outage response. It enters the model before line 17 does. However its coefficient is overtaken by that of line 17 as they increase towards the least squares solution, giving the correct final identification result.




\subsection{Indistinguishable Line Outages}
\label{sec:diagnosability}
As seen from the signature map of Fig. \ref{fig:feature_map}, some outages  create highly similar responses from the system, i.e., $F_i \approx F_j$ for some $i,j \in \mathcal{L}$. In general, this ambiguity problem is commonly encountered in realistic systems \cite{Wu2015}. One reason is that some line outages do indeed create similar responses due to a combination of topological positions and pre-outage power flow carried. On the other hand, a limited PMU deployment might mean distinctive signatures of some outages are not observable. Intuitively, the second situation is more pronounced as the PMU budget decreases. It is also well-known that with a group of highly correlated predictors, the lasso formulation of (\ref{eqn:lasso_formulation}) tends to select one from the group and does not care which one to select \cite{zou2005regularization}. In the extreme case where $F_{i} = F_{j}$ for some $i,j \in \mathcal{L}$ and $\boldsymbol{\hat{\beta}}$ is the lasso solution, it can be shown that $\hat{\beta}_{i}\hat{\beta}_{j} \ge 0 $ and $\boldsymbol{\hat{\beta}}^*$ is another solution of (\ref{eqn:lasso_formulation}) where
\begin{align}
\hat{\beta}^*_k = 
\begin{cases}
\hat{\beta}_k \,,  k \ne i, k \ne j \\
(\hat{\beta}^*_{i}+\hat{\beta}^*_{j}) s \,, k=i \\
(\hat{\beta}^*_{i}+\hat{\beta}^*_{j}) (1-s) \,, k=j \,,
\end{cases}
\end{align}
for some $s \in [0, 1]$ and $k \in \mathcal{L}$. Therefore, lasso might not have a unique solution when predictors are highly correlated.

In the context of our identification problem, the true outage line, e.g., $i$, might not be correctly identified if $F_i \approx F_j$, and equivalently their correlation is close to 1, 
$$
\operatorname{corr}(p_i F_i, p_j F_j) = \operatorname{corr}(F_i, F_j) \approx 1 \,,
$$
for some $i, j \in \mathcal{L}$. To address this ambiguity problem, we propose to group transmission lines into minimal diagnosable clusters (MDCs). Each MDC contains lines which, given a fixed PMU placement, produce responses that our lasso formulation could not distinguish with a high probability. Concretely, we define MDC as a group of lines whose observable outage effects have pairwise correlations higher than a pre-defined threshold $\rho^*$. Therefore, the MDC of line $i$ is 
\begin{equation}
\label{eqn:mdc_threshold}
g_i = \{j:\text{corr}(F_{i}, F_{j}) \ge \rho^*\} \,,
\end{equation}
for $j \in \mathcal{L}\setminus i$. The collection of MDCs for all transmission lines is
\begin{equation}
\label{eqn:mdc_collection}
G_F = \{g_1, g_2, \dots, g_L\} \,.
\end{equation}
Also, the diagnosability of a system with given PMU locations can be characterized by the proportion of single-element MDCs,
$$
V(\rho^*) = \left(\sum_{i=1}^{L} \mathbf{1}(|g_i| = 1)\right)/L \,,
$$
where $\mathbf{1}(\cdot)$ is indicator function and $|g|$ counts the number of elements in the set $g$. 

Intuitively, a smaller $\rho^*$ corresponds to a more relaxed correlation requirement to enter the MDC, therefore in general decreases $V(\rho^*)$ and vice versa. With MDCs constructed offline, they can augment the lasso solution in real-time outage identification. Suppose $L_o = \{l_i^*, i=1, \dots, a\}$ are identified by lasso as outage lines. The augmented solution set would be 
\begin{equation}
\label{eqn:mdc_augmentation}
L_o^* = \{g_{l_1^*}\cup \cdots \cup  g_{l_a^*}\} \,.
\end{equation}
With MDC augmentation, outage identification accuracy is improved, however, potentially at the expense of identification precision. The trade-off is controlled through the correlation threshold. The effect of the threshold and accuracy-precision trade-off is investigated further in simulation study of Section \ref{sec:simulation}. 

\begin{figure}[t]
\centering
\includegraphics[width=1\linewidth]{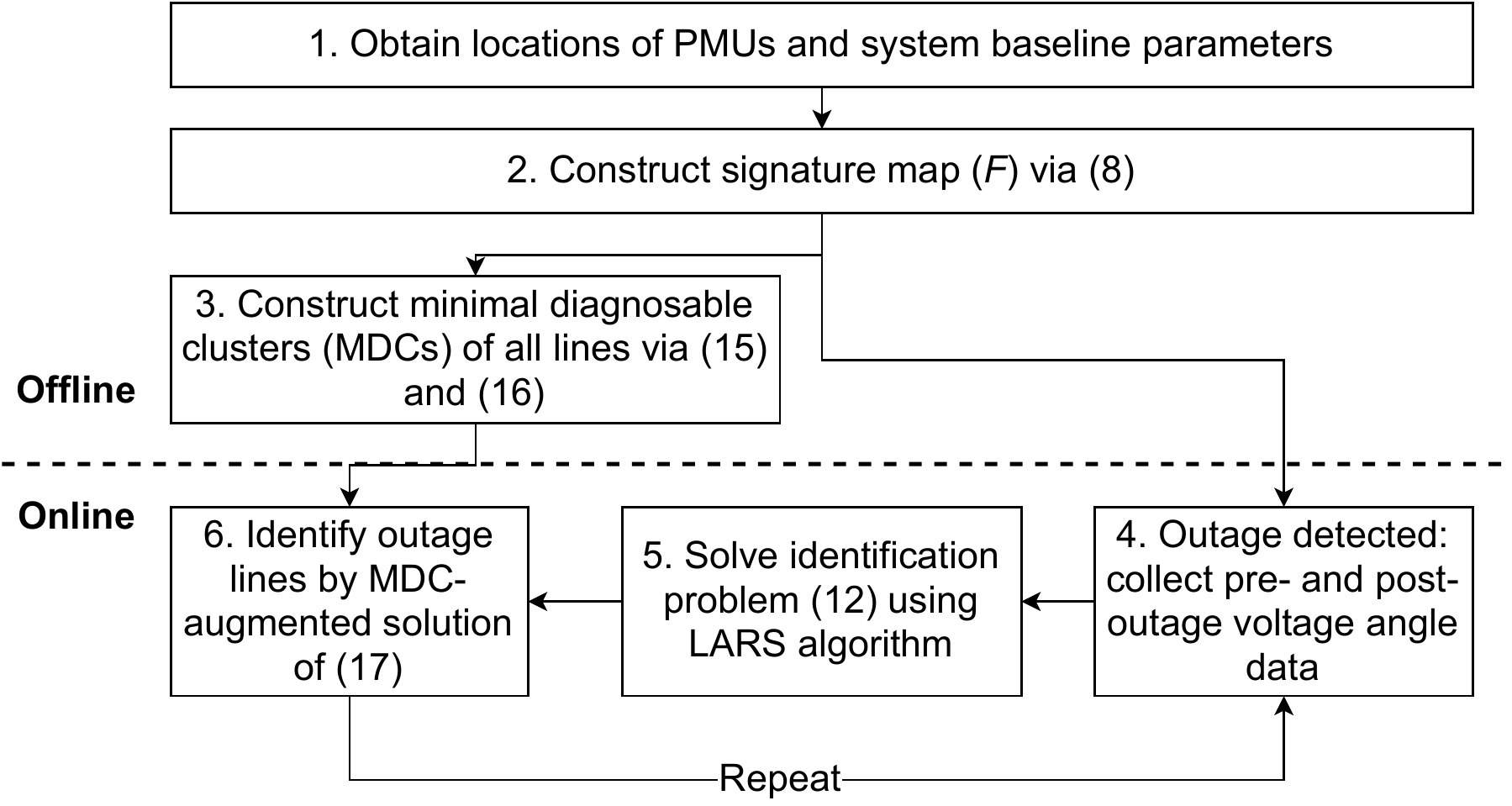}
\caption{\textit{Framework of the proposed line outage identification scheme. Preparation steps one to three can be performed offline while outage identification steps four to six can be carried out during real-time monitoring operations.}}
\label{fig:flow_chart}
\end{figure}

To end this section, we summarize the proposed identification scheme in Fig. \ref{fig:flow_chart}. The scheme is split into an offline and online part. Preparation work of step one to three could be done offline since they only require quasi-steady state information and baseline system parameters. Once the signature map and MDCs are constructed, they could be used in real-time monitoring operation as in step four to six.

The idea of constructing expected angle change based on power injection to identify outage lines is not new \cite{Tate2008,Wu2015,Costilla-Enriquez2019}. Separately, authors in \cite{Zhu2012} and \cite{Chen2014} have also formulated outage identification as a sparse vector recovery problem. However our work is different in the following aspects: 1) All except \cite{Costilla-Enriquez2019} have relied on the simplified DC power flow model by assuming a flat voltage profile and approximately identical phase angles. Our signature map is derived from the AC power flow model, better reflecting the heterogeneous operating condition of power systems. 2) All except \cite{Wu2015} do not consider the impact of indistinguishable outage events on identification performance. Whereas Wu \etal conduct online search for indistinguishable outage locations, our MDCs are constructed in advance and incur no extra computation during real-time identification. 3) Enriqurez \etal uses both voltage and current phasor for identification while our method only requires voltage measurements \cite{Costilla-Enriquez2019}. Performance of a sparse regression-based \cite{Zhu2012} and an AC power flow-based method \cite{Costilla-Enriquez2019} are compared in our simulation study.


\section{Simulation Study}
\label{sec:simulation}

\subsection{Simulation Setting}
We test our identification scheme on IEEE 39-bus New England test system \cite{athay1979practical}. System transient responses following an outage are simulated using the open-source simulation package COSMIC \cite{Song2016} in which a third-order machine model and AC power flow model are used. We assume that the sampling frequency of a PMU is 30 samples per second. The system loads are varied by a random percentage between -5\% and 5\% from the base-line values for each simulation run. The total duration of a run is 10 seconds; the outage takes place at the 3rd second. Pre- and post-outage voltage phase angles are obtained at the 1st and 10th second. Artificial noise is added to all sampled angle data, $\Delta \boldsymbol{\theta}$, to account for system and measurement noise. They are drawn from a Gaussian distribution with mean $\mathbf{0}$ and standard deviation of $5\%$ of the pre-outage $\Delta\boldsymbol{\theta}$ on respective buses. 

Simulated single-line outages include line 1 to 36 except line 21 as it creates two islands. Double-line outages include 100 random pairs of lines from line 1 to 46 that does not create separate islands. Given a list of identified and true outage lines, $L_o$ and $L_{true}$, identification performance is assessed by Accuracy (A), 
\begin{equation}
\label{eqn:accuracy}
A(L_o, L_{true}, a):=\frac{\sum_i \mathbf{1}(\left| L_{o,i} \cap L_{true, i} \right| = a)}{\left| L_{true} \right|} \,,
\end{equation}
Therefore, the accuracy of single-line outage identification of a scheme is $A(L_o, L_{true}, 1)$. Similarly, the ``half-correct'' and ``all-correct'' accuracy of double-line outage is $A(L_o, L_{true}, 1)$ and $A(L_o, L_{true}, 2)$. Accuracy with MDC augmentation for each scenario is obtained by replacing $L_o$ with $L_o^*$, the augmented set defined in (\ref{eqn:mdc_augmentation}).

\subsection{Illustrative Outage Identification Example}  

\begin{figure}[t]
\centering
\includegraphics[width=1\linewidth]{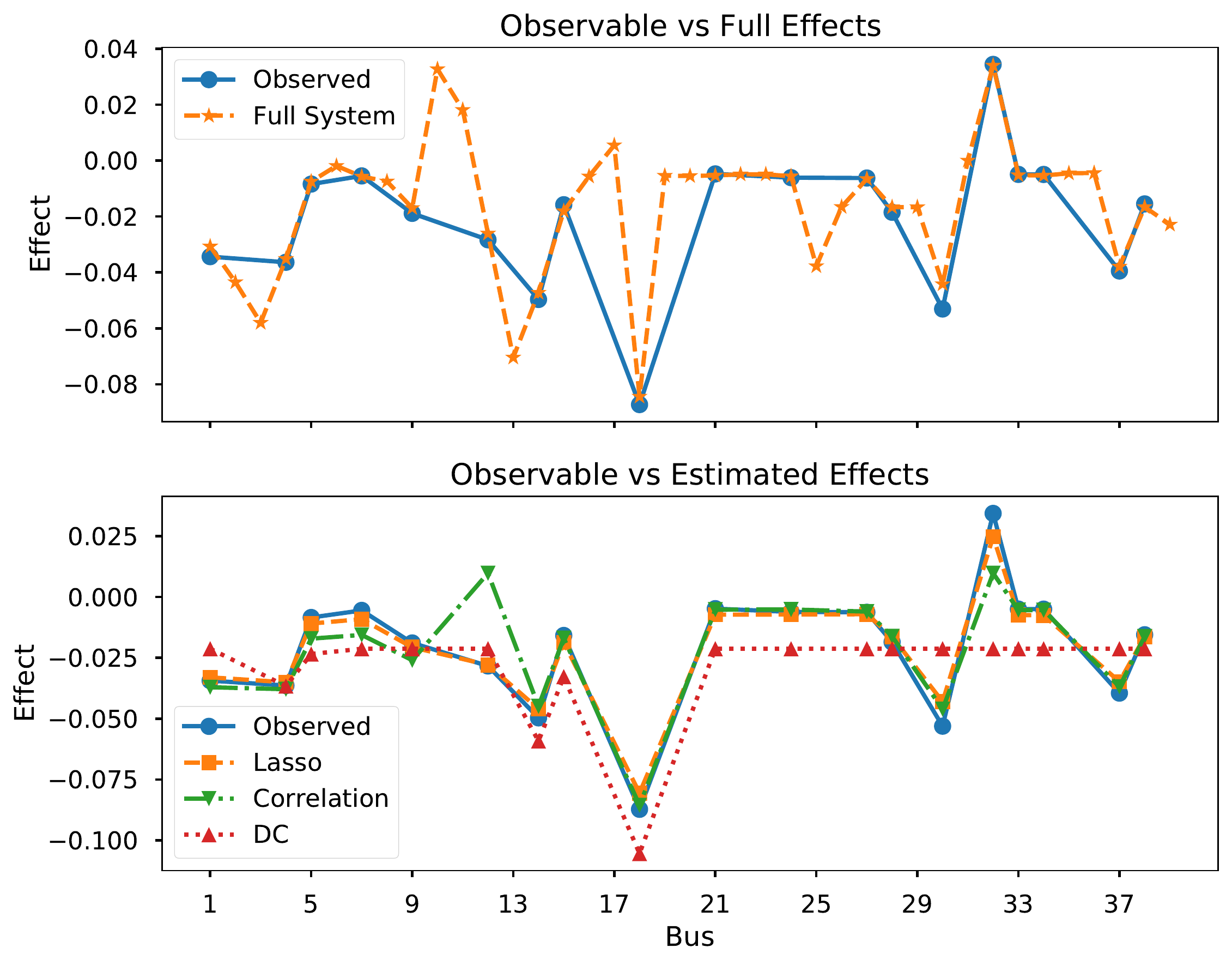}
\caption{\textit{Full, observed, and estimated outage impact on bus voltage phase angles after a double-line outage at line 17 and 25. 19 out of 39 buses are equipped with PMUs. The top figure shows observed noisy data with true and complete system states. The bottom figure compares the estimated phase angles changes from three methods against the observed states.}}
\label{fig:estimation_comparison}
\end{figure}

Using the same example of a double-line outage at line 17 and 25, Fig. \ref{fig:estimation_comparison} demonstrates limited observability (top) and the estimation of $\Delta\boldsymbol{\theta}$ by each method (bottom). The angle change estimation is obtained using recovered power transfer coefficient $\boldsymbol{\hat{\beta}}$, i.e., $\Delta\boldsymbol{\hat{\theta}} = F \boldsymbol{\hat{\beta}}$. Limited deployment of PMUs means some bus angles are not observed. This is illustrated in the top figure where some signatures of the outage are not missed. If unobserved locations contain all the distinctive signatures of that outage, distinguishing it from the others would be challenging. Therefore, characterizing and exploiting line diagnosabilities through MDCs are necessary to overcome this challenge. 

The bottom figure shows a comparison of $\Delta\boldsymbol{\hat{\theta}}$ by three methods under comparison. Columns of $F$ corresponding to the outage lines identified by each method are used. AC power flow-based methods are clearly better at reconstructing the angle changes than the DC one. Notice that the DC estimation has more ``flat'' angles than the other two, thus fewer details to distinguish it from other outages. While variable selection accuracy rather than estimation accuracy is our focus, this figure nevertheless demonstrates the superior performance of AC power flow model at capturing a more nuanced outage impact.

\subsection{Average Identification Performance}  
Average performance of each identification scheme is reported based on 200 simulation runs over all the single- and double-line outages. Random noises and PMU placements of a 25\% or 50\% PMU coverage are used in each run. Two existing methods are compared, namely ``DC'' for the DC power flow-based method in \cite{Zhu2012} and ``Corr'' for the AC power flow-based method in \cite{Costilla-Enriquez2019}. Performance gain with MDC augmentation of (\ref{eqn:mdc_augmentation}) is also reported under the name ``...+MDC''.

\subsubsection{Single-line outage}  
\begin{figure}[htpb]
\centering
\includegraphics[width=1\linewidth]{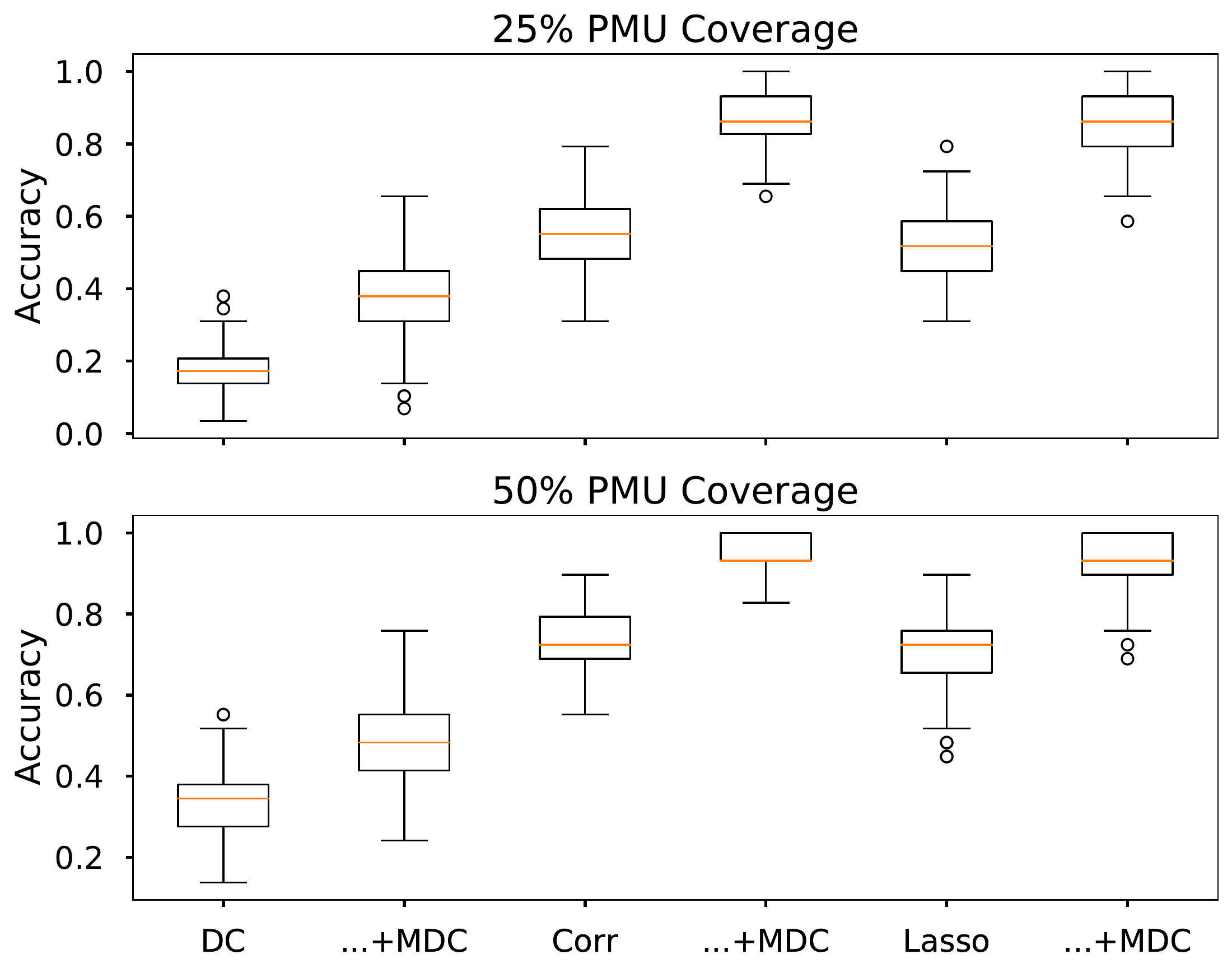}
\caption{\textit{Box-plots of single-line outage identification results for DC-based, correlation-based, and the proposed method. Results are based on 200 random simulation runs under a 25\% (top) and 50\% (bottom) PMU coverage in the New England 39-bus system. Each method has two sets of results: accuracy of the original identification and of that augmented with MDCs.}}
\label{fig:single_outage}
\end{figure}

Fig. \ref{fig:single_outage} shows the identification results for single-line outage. With or without MDC augmentation, correlation-based method and the proposed method consistently outperform the DC-based method in both cases of PMU coverage. The former two methods are roughly always 40\% more accurate than the DC-based method. Correlation-based method achieves almost identical result with the proposed method regardless of MDC augmentation or PMU coverage. This is expected since the proposed method identifies the first variable as the one most correlated with the response, an identical procedure as the correlation-based method.

When PMU coverage is increased from 25\% to 50\%, improvements in identification accuracy across all methods are observed as expected. Under a 25\% coverage, the proposed method is 52\% and 86\% accurate (median), without and with MDC augmentation. With a 50\% coverage, the performance is 72\% and 93\%. Lastly, augmenting original solution with their MDCs improves accuracy across methods and PMU coverage. Roughly speaking, MDC augmentation improves accuracy by 30\% for the 25\% coverage and 20\% for the 50\% coverage. Notably, the two AC-based methods reach 93\% identification accuracy under a 50\% coverage with MDCs. The decrease in accuracy improvement for better observed system might be because they tend to have more distinguishable outages. 

\subsubsection{Double-line outage} 
\begin{figure}[htpb]
\centering
\includegraphics[width=1\linewidth]{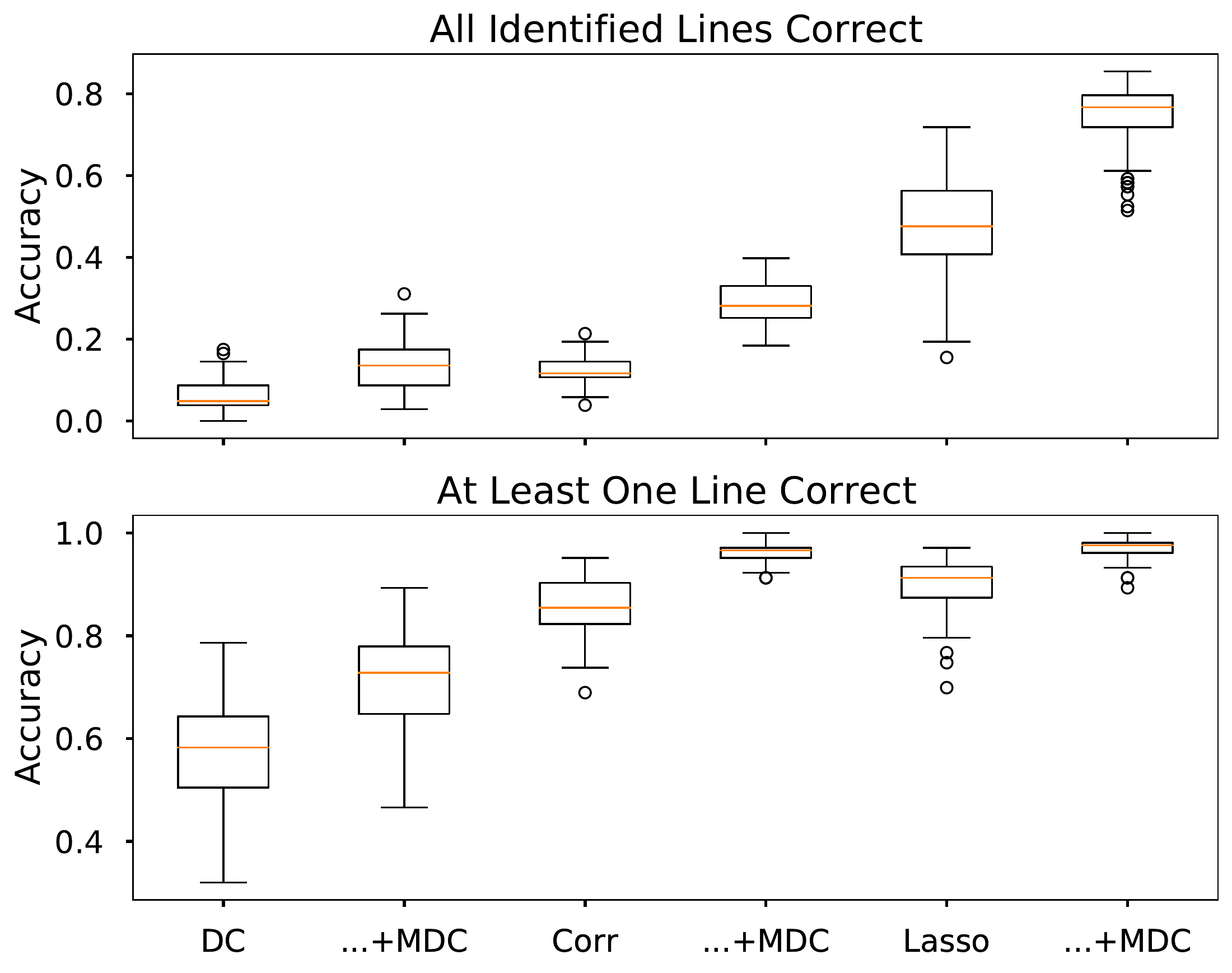}
\caption{\textit{Box-plots of double-line outage identification results for DC-based, correlation-based, and the proposed method. ``All correct'' (top) and ``half correct'' (bottom) results are based on 200 random simulation runs under a 50\% PMU coverage in the New England 39-bus system. Each method has two sets of results: accuracy of the original identification and of that augmented with MDCs.}}
\label{fig:double_outage}
\end{figure}

Fig. \ref{fig:double_outage} shows the identification results for double-line outage. The proposed method consistently outperforms the other two methods, especially in the ``all correct" case. DC-based method performs worst in both categories. The correlation-based method is not as accurate beyond the identification of the first line. The reason might be that the proposed formulation treats $\boldsymbol{p}'$ as an unknown vector. It is systematically estimated from data by lasso. However the correlation-based method treats it as a fixed vector of line reactance. Inaccuracy in the model might then lead to inaccurate identification of multiple outage lines. Again, augmenting solutions with MDCs improve accuracy for all methods, especially in the ``all correct'' category. Overall, the proposed method with MDC augmentation (Lasso+MDC) achieves the best performance. It can identify 80\% of the simulated double-line outages under a 50\% PMU coverage.

\subsubsection{Effect of minimal diagnosable cluster}  

\begin{table}[htpb]
\caption{Impact of Minimal Diagnosable Cluster Threshold on Identification Precision-Accuracy Trade-off Using Lasso+MDC}
\label{tab:impact_mdc_threshold}
\centering
\begin{tabular}{llll}
\hline
\hline
Threshold ($\rho^*$)  & Single-element MDC (\%) & Single-line & Double-line \\
\hline
0.80 & 0.34 (0.06) & 0.94 (0.06) & 0.69 (0.08) \\
0.84 & 0.42 (0.06) & 0.94 (0.05) & 0.69 (0.07) \\
0.88 & 0.49 (0.07) & 0.95 (0.05) & 0.69 (0.08) \\
0.93 & 0.55 (0.06) & 0.93 (0.06) & 0.67 (0.09) \\
0.95 & 0.58 (0.06) & 0.93 (0.07) & 0.66 (0.09) \\
0.98 & 0.62 (0.06) & 0.92 (0.06) & 0.66 (0.09) \\
0.99 & 0.68 (0.07) & 0.89 (0.07) & 0.61 (0.09) \\
\hline 
\end{tabular}
\end{table}
Table \ref{tab:impact_mdc_threshold} shows the trade-off between identification precision and accuracy by varying the MDC threshold $\rho^*$ in (\ref{eqn:mdc_threshold}). As expected, the proportion of single-element MDC, $V(\rho^*)$, increases as the threshold approaches 1. In particular, a threshold of $\rho^* = 0.8$ leads to 34\% MDCs with a single element. That proportion increases to 68\% when $\rho^* = 0.99$. The good news is, a tighter requirement on MDC does not lead to much decrease in single- and double-line outage identification accuracy using the proposed method, from 94\% to 89\% and 69\% to 61\%, respectively. 

Therefore, augmenting solution with MDCs could substantially improve identification accuracy while sacrificing a moderate amount of identification precision. The result also suggests that recognizing the most highly correlated line outages, e.g., setting $\rho^* \ge 0.95$, is enough to reap the benefit of MDC augmentation. Nevertheless, there is a trade-off between accuracy and precision. The threshold could be determined in conjunction with decision-makers' other considerations, e.g., resources available or criticality of the system. 

\subsubsection{Effect of measurement noise} 

\begin{figure}[htpb]
\centering
\includegraphics[width=1\linewidth]{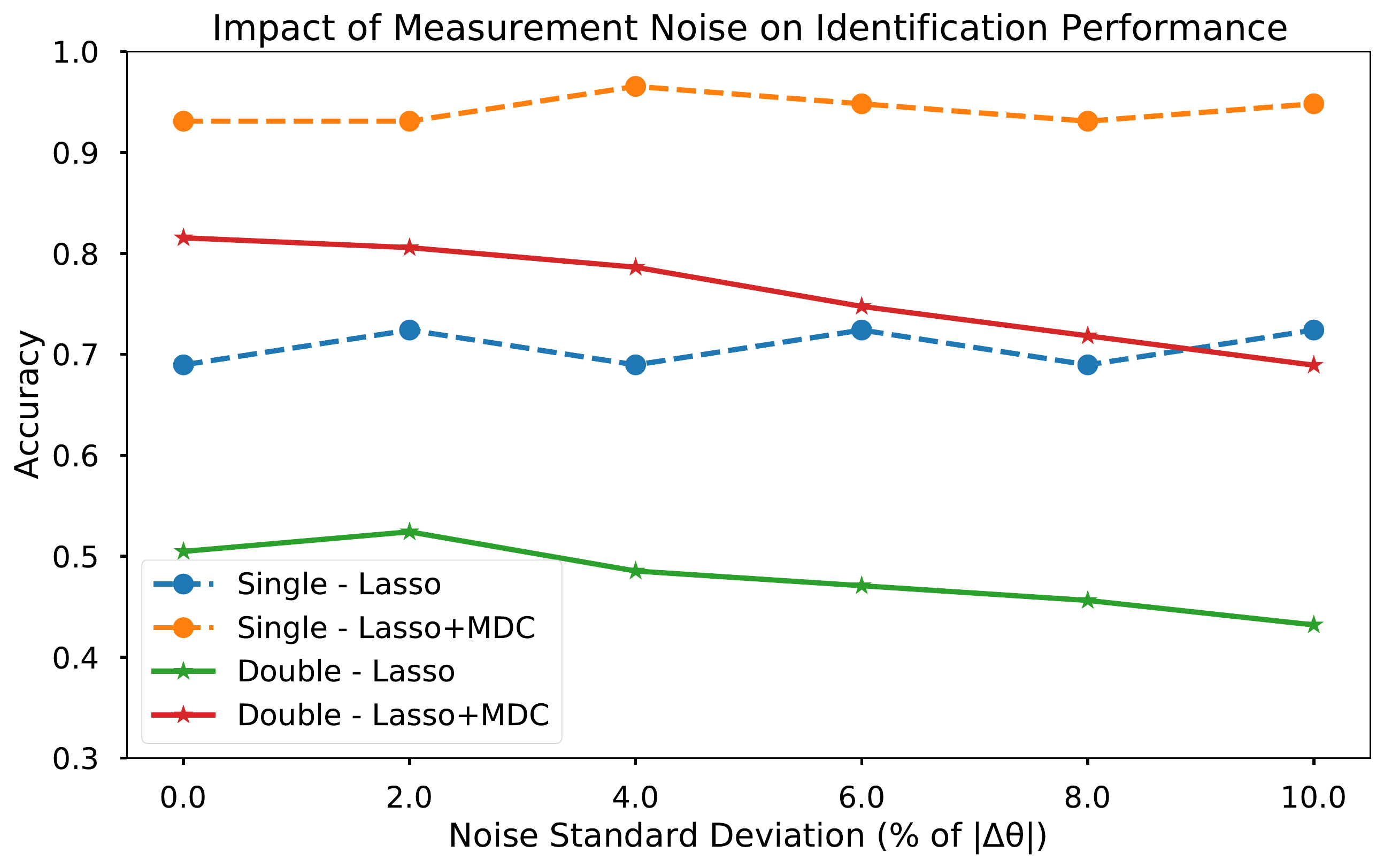}
\caption{\textit{Impact of measurement noise on identification performance of the proposed method. Performance using data with noise standard deviation varying from 0\% to 10\% of $|\Delta\boldsymbol{\theta}|$ is reported by median accuracy of single- and double-line outages using Lasso and Lasso+MDC. }}
\label{fig:impact_noise}
\end{figure}

We also report the performance of the proposed method with respect to measurement noise in Fig. \ref{fig:impact_noise}. The performance is largely robust to measurement noise. The accuracy for single-line identification shows no clear difference as the noise level increases. There is a moderate decrease in accuracy for double-line outage identification as the noise level increases to 10\%. Lasso formulation is known to be robust to noise \cite{hastie2020best}. This is corroborated by results from our simulation study. 

\section{Conclusion}
\label{sec:conclusion}
In this paper, we propose a novel framework of real-time multiple-line outage identification with limited PMU deployment. AC power flow model is utilized to construct a signature map that encodes voltage phase angle signatures of each line outage. Identification is then formulated into an underdetermined sparse regression problem solved by lasso. Minimal diagnosable clusters are proposed to further improve identification accuracy. 
Single-line and double-line outages simulated on the New England 39-bus system with 25\% and 50\% PMU deployment are used to study the proposed method's performance. The proposed method is shown to have better identification accuracy under all simulation settings, especially for double-line outages. The robustness of the method is also demonstrated using varying levels of noisy data. Finally, we have also shown the merit of exploiting line diagnosability through minimal diagnosable cluster which significantly improves identification accuracy by trading off a small amount of precision. 

We did not however consider the problem of post-outage system parameter recovery in this work. In general, online updating of system parameters under a partial observability remains a challenging and important task that is worth investigating. Also, the diagnosability of line outage is intimately related to where the limited number of PMUs are placed. We intend to study the optimal PMU placement problem for line outage identification in our future research.


\section*{Acknowledgment}
The authors thank Dr. Chao Zhai and Dr. Yuchen Shi for insightful discussions during the inception of this work. This research was conducted at the Singapore-ETH Centre, which was established collaboratively between ETH Zurich and the National Research Foundation Singapore. This research is supported by the National Research Foundation, Prime Minister’s Office, Singapore under its Campus for Research Excellence and Technological Enterprise (CREATE) programme.

\ifCLASSOPTIONcaptionsoff
  \newpage
\fi



%
\begin{IEEEbiography}{Xiaozhou Yang}
received the B.S. degree in industrial and systems engineering from the National University of Singapore, Singapore, in 2017. 
He is currently a postdoctoral researcher at the Future Resilient Systems program of Singapore-ETH Centre. 
His research interests include advanced data analytics in power system condition monitoring, real-time outage detection, and identification.
\end{IEEEbiography}

\begin{IEEEbiography}{Nan Chen}
received the B.S. degree in automation from Tsinghua University, Beijing, China, in 2006, the M.S. degree in computer science in 2009, and the M.S. degree in statistics and the Ph.D. degree in industrial engineering from the University of Wisconsin-Madison, Madison, WI, USA, both in 2010.
He is currently an Associate Professor with the Department of Industrial Systems Engineering and Management, National University of Singapore, Singapore. His research interests include statistical modeling and surveillance of engineering systems, simulation modeling design, condition monitoring, and degradation modeling.
\end{IEEEbiography}




\end{document}